\documentclass[preprint,showpacs,preprintnumbers,amsmath,amssymb]{revtex4}
\usepackage{graphicx}

\begin{document}
\title{\large{Polarization Rotation, Switching and E-T phase diagrams of BaTiO$_3$: A Molecular Dynamics Study}}
\author{Jaita Paul$^1$, Takeshi Nishimatsu,$^2$, Yoshiyuki 
Kawazoe$^2$ and Umesh V. Waghmare$^1$}
\affiliation{$^1$TSU, Jawaharlal Nehru Centre for 
Advanced Scientific Research, Jakkur PO
Bangalore 560 064 India\\
$^2$Institute for Materials Research, Tohoku University, Sendai 980-8577, 
Japan}

\begin{abstract}
We use molecular dynamics simulations to understand the mechanisms of 
polarization switching in ferroelectric BaTiO$_3$ achieved with external 
electric field. For tetragonal and orthorhombic ferroelectric phases, we 
determine the switching paths, and show that  polarization rotation 
through intermediate monoclinic phases (a) facilitates switching at low 
fields (b) is responsible for a sharp anisotropy in polarization switching. 
We develop understanding of this through determination of detailed electric 
field-temperature phase diagrams, that are fundamental to 
technological applications based on electromechanical and switching response
of ferroelectrics.

\end{abstract}
\maketitle

Ferroelectric (FE) materials exhibit spontaneous electric polarization 
whose magnitude and direction depend 
sensitively on temperature, pressure and electric field\cite{dawber}.
They are technologically very important because of two of their properties:
(a) large electromechanical coupling that is exploited in their applications
as sensors and actuators in micro-electro-mechanical systems \cite{dawber}, and (b) 
switchability of their polarization from one state to another with applied
field, that makes them useful in non-volatile memory devices like 
ferroelectric random access memories (FeRAMs)\cite{scott}. Fundamental
understanding of the mechanisms responsible for these properties in
the bulk crystals is essential to development of ferroelectrics with 
improved properties and their use in nano-scale 
devices.

Direction of spontaneous polarization is typically along a crystallographic
direction. For example, tetragonal, orthorhombic and rhombohedral ferroelectric
states are characterized by polarization along [001], [110] and [111] 
directions, respectively. Polarization rotation\cite{fu-cohen} from a 
rhombohedral state towards a tetragonal one through an intermediate monoclinic phase
was shown to be responsible for an ultrahigh piezoelectric coupling 
observed experimentally\cite{park1}. Monoclinic phases, characterized by
the polarization vector ({\bf{P}}) along low symmetry directions, were carefully
analyzed and shown to be relevant to the large piezoelectricity
in 92\%PbZn$_{1/3}$Nb$_{2/3}$O$_3$-8\%PbTiO$_3$ \cite{noheda1} 
and also found in another technologically important ferroelectric,
Pb(Zr$_x$Ti$_{1-x}$)O$_3$\cite{pandey}.

Interestingly, piezoelectric response of BaTiO$_3$ (a simple, classic ferroelectric)
was shown to be enhanced significantly by crystallographically 
engineering single crystals through applied DC bias\cite{wada, park2}. 
This involves nonlinear response of BaTiO$_3$ and is linked with
electric field induced structural phase transitions\cite{park1, paik, noheda1}
in ferroelectrics. Polarization switching, crucial to memory applications,
is another closely related nonlinear phenomenon whose microscopic
mechanism is yet to be uncovered. Understanding of these nonlinear phenomena
within a single picture is facilitated by knowledge of electric
field-temperature phase diagrams of a ferroelectric.

Using phenomenological 
Landau-Ginzburg-Devonshire theory, 
electric field-temperature ($E-T$) phase 
diagrams have been studied for single crystal BaTiO$_3$, showing
evidence for monoclinic phases\cite{bell}. Similarly, it was shown in
general by Vanderbilt and Cohen \cite{vand} that the extension of the sixth order 
Devonshire theory to eighth and twelfth orders in polarization as the 
order parameter was necessary to explain the presence of stable monoclinic and triclinic 
phases, respectively.

Here, we present molecular dynamics simulations carried out for 
a comprehensive investigation of polarization switching
in BaTiO$_3$ through determination of (a) its temperature dependence,
(b) paths followed during switching with field along different directions,
and (c) detailed electric field-temperature phase diagrams. This permits
identification of different monoclinic phases that are stabilized 
as a function of electric field and temperature, and are relevant to both ultra-high
piezoelectric response and polarization switching properties.

We use a first-principles effective Hamiltonian in classical molecular 
dynamics simulations that (1) capture most nonlinearities by construction,
and (2) include thermal fluctuations in polarization and strain. 
A similar approach has been used earlier\cite{bellaiche} in finding
polarization paths as a function of applied field at a fixed temperature.
An effective Hamiltonian $H_{\rm eff}$ is a Taylor series expansion 
of the energy surface around the high symmetry cubic perovskite structure, written in 
terms of the low energy local degrees of freedoms (DoFs), three acoustic phonon modes
(which capture the physics of inhomogeneous strain) and three low energy (soft)
optical phonon modes (which give local dipole moments).  
In order to include effects of an external electric field on the system, a term 
$-Z^*\sum_{\bm{R}}\bm{\mathcal{E}}\!\cdot\!\bm{u}(\bm{R})$
where $Z^*$ and $\bm{u}(\bm{R})$ are the Born effective charge and
displacement associated with the soft-mode variable of the
unit cell at $\bm{R}$,
is added to $H_{\rm eff}$.
We use parameters in $H_{\rm eff}$ as determined in Refs \cite{zhong} and \cite{rabe}
from first-principles density functional theory calculations with local density approximation (LDA). 
Use of LDA causes an underestimation of the 
lattice parameter, hence a negative pressure of -5 GPa is applied to all systems simulated \cite{rabe} here. 

Mixed space molecular dynamics\cite{umesh1, umesh2} (MD)
is used here to determine finite temperature 
properties of H$_{\rm eff}$ with periodic boundary conditions. 
The FERAM code used in our simulations is described in detail in Ref\cite{takeshi} 
and the program can be downloaded from 
\url{http://lotto.sourceforge.net/feram}.
In order to simulate polarization hysteresis, the temperature is first increased or
decreased in steps of $\sim$30 K to achieve an equilibrium at the temperature 
of the hysteresis simulation.  Then, at fixed temperature, the electric field is switched on 
from zero to a positive value E$_{max}$, from which it is decreased to -E$_{max}$ 
in steps of about 10 kV/cm to 100 kV/cm. Subsequently,
it is again increased finally to E$_{max}$ to obtain a complete hysteresis loop. 
At each temperature (and electric field), the system is thermalized for initial 70,000 
timesteps and then averaging is performed over the next 
30,000 timesteps, amounting to a simulation period of 0.2 ns. 

For the high temperature cubic paraelectric phase, we simulate response to
electric field applied along [001] direction (symmetry would give the same
 response to fields along other directions). 
The Langevin function
\begin{eqnarray}
\centering
\centerline{$L(E) = k_1[coth\frac{k_2E}{k_BT}-\frac{k_BT}{k_2E}]$}
\end{eqnarray}
describes fairly well the P(E) curves of the paraelectric phase (see Fig 1a and
fitting parameters $k_1$ and $k_2 $ in Table 1). 
While $k_1$, which relates to the saturated polarization remains roughly 
constant, $k_2$, which is proportional to the dielectric constant of the paraelectric phase,
diverges as T approaches T$_c(T\leftrightarrow C)$. Almost perfect fits to P(E) curves 
of the paraelectric phase with just two parameters and their expected temperature dependence
validates our MD simulations to study further these nonlinear hysteresis properties of the
polar phases.

For polar phases, we simulate response to field applied (a) along 
the respective polar axis and (b) along other high symmetry crystallographic 
directions (eg., along [101] direction of the tetragonal phase where {\bf P} points 
along [001]). Generally, the magnitude of field at which switching occurs, $\mathcal{E}_c$,
reduces as the temperature increases upto the transition to higher temperature
phase (see results for tetragonal and rhombohedral phases in Fig 1b and 1c).

\begin{center}
\begin{table}
\caption{
Parameters in the Langevin function used in fitting P(E) data of 
the paraelectric phase at different temperatures.}
\begin{tabular}{|c|c|c|c|}
\hline
T (K) & $k_1$ (C/m$^{2}$) & $k_2$ (Cm$\times$10$^{-5}$) \\ \hline \hline
400&0.12&25.4 \\ \hline
370&0.12&41.5 \\ \hline
340&0.11&101.8 \\ \hline
\end{tabular}
\end{table}
\end{center}
We now focus on the direction-dependence of polarization switching in rhombohedral 
phase. We find that the coercive field $\mathcal{E}_c$ when applied along [001] direction is 
almost a factor of 5 smaller than that when applied along [111] direction. While the
reduction in $\mathcal{E}_c$ with T can be readily understood through the concept of 
lowering  of Landau free energy barrier near the transition,
a strong dependence of $\mathcal{E}_c$ on its direction is related to the fact that transition
state itself could be different.

\begin{figure}[htbp]
\centering
\includegraphics[scale=0.3]{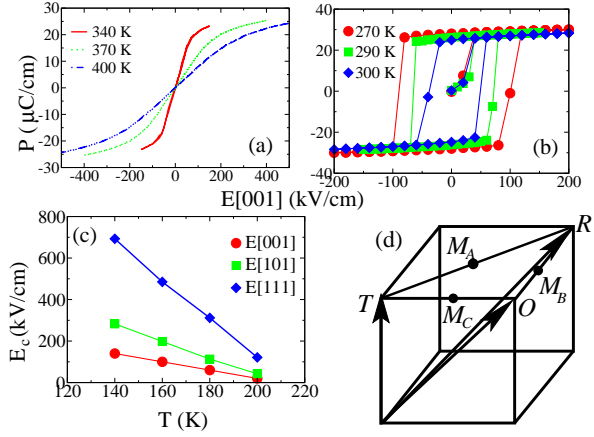}
\caption{(Color online) Average value of polarization P$_z$ as a function of electric field 
E for (a) cubic paraelectric and (b) tetragonal ferroelectric 
phases when electric field is applied along [001] direction. (c) Coercive field $\mathcal{E}_c$ as a function
of temperature for fields along [111], [101] and [001] directions respectively,
applied to rhombohedral phase. (d) Cohen's Cube depicting the high symmetry tetragonal (T),
orthorhombic (O) and rhombohedral (R) phases and
monoclinic M$_A$, M$_B$ and M$_C$ phases.}
\end{figure}

In order to investigate the transition states relevant to switching,
we examine the paths followed by the system during
polarization switching. When E is parallel to {\bf{P}},
the path is linear ({\bf{P}} to {\bf{-P}}) by symmetry. We use the following 
nomenclature to describe the different phases
encountered along the polarization paths (Fig. 2): 
T, O and R denote the tetragonal, orthorhombic and 
rhombohedral phases respectively, and, 
$M_A$, $M_B$ and $M_C$ denote the three different 
monoclinic phases as defined in Ref.\cite{vand}. 
In the $M_C$ phase, {\bf{P}} points along [0uv] (u$\not=$v), 
and in the $M_A$ and $M_B$ phases {\bf{P}} points 
along [uuv], with u$<$v and u$>$v, respectively. The monoclinic
phases are also illustrated with the Cohen's cube\cite{fu-cohen}
(see Fig. 1d).

\begin{figure}[htbp]
\centering
\includegraphics[scale=0.3]{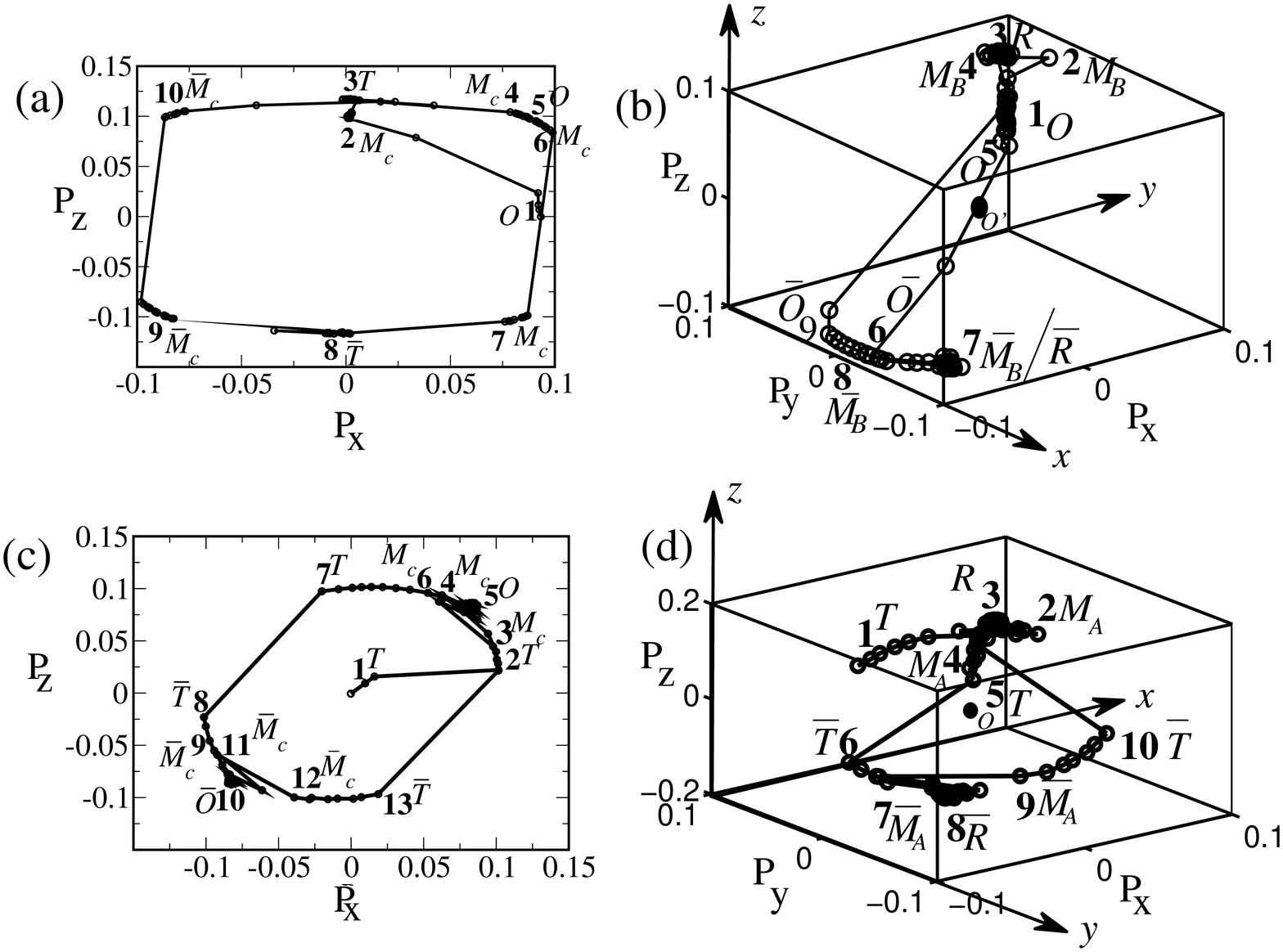}
\caption{Paths of polarization traced during switching when 
field is applied along (a) [001] and (b) [111] directions to 
orthorhombic phase at 240 K and (c) [101] and (d) [111] directions to tetragonal
phase at 290 K. The numbers mark the sequence of phases (nomenclature in the text) 
followed by {\bf{P}}.}
\end{figure}

With field along [001] and [111] directions, switching in the
orthorhombic phase (at 240 K) occurs through monoclinic phases $M_C$ and $M_B$ 
that appear along the path at 25 kV/cm and 15$\sqrt{3}$ kV/cm,
respectively. The transition from orthorhombic to the
respective monoclinic phases is marked by (1) a jump in the
polarization path from point 1 to 2 in Fig. 2a and 2b, and 
(2) a change in the slope of the curve of eigenvalues $\eta$ of
strain tensor as a function of electric field (Fig. 3a and 3b). 
The second crossover ($M_C$$\rightarrow$$T$ and $M_B$$\rightarrow$$R$) 
manifests as a change in the slope of $\eta$ at $\sim$ 60 kV/cm (Fig. 3a)
and $\sim$ 50 kV/cm (Fig. 3b), respectively.
The sequence of phases, O$\rightarrow$$M_C$$\rightarrow$T 
(Fig. 2a) and O$\rightarrow$$M_B$$\rightarrow$R (Fig. 2b) can be readily rationalized
through continuous paths along the Cohen's cube (see Fig. 1d). 

Along the polarization paths for tetragonal phase (at 290 K), 
the monoclinic phases $M_C$ and $M_A$ emerge when fields 
point along [101] (Fig. 2c) and [111] (Fig. 2d)
directions, respectively. The transitions from tetragonal to the monoclinic 
phases are evident in the changing slopes at $\sim$ 40 kV/cm of $\eta$ (Fig. 3c and 3d).
Only at higher fields $\sim$ 100 kV/cm, close to E$_{max}$, 
the phases become orthorhombic and rhombohedral, respectively. The sequence of
phases T$\rightarrow$$M_C$$\rightarrow$O and T$\rightarrow$$M_A$$\rightarrow$R
also follows from Cohen's cube (Fig. 1d).

\begin{figure}[htbp]
\centering
\includegraphics[scale=0.27,angle=-90]{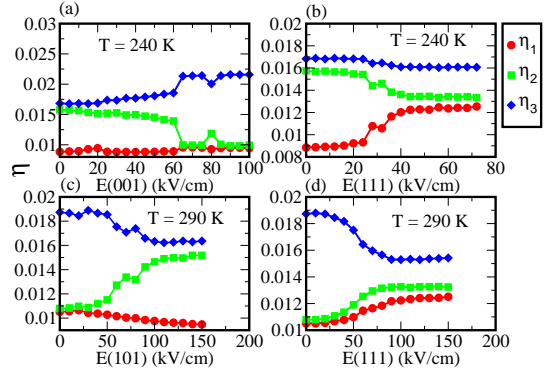}
\caption{(Color online) Eigenvalues $\eta$ of strain tensor as a function of 
field applied along (a) [001] (b) [111] directions to
the orthorhombic phase at 240 K and (c) [101] (d) [111] directions to 
the tetragonal phase at 290 K.}
\end{figure}

At a field of 25 kV/cm along [001] direction of the orthorhombic phase,
there occurs a jump (from point 6 to 7 in Fig. 2a) along the path of switching 
which corresponds to a rotation from $M_C$ to $\bar{M_C}$. 
Similarly when the field is 15$\sqrt{3}$ kV/cm along [111] direction,
a jump from point 5 to 6 (see Fig 2b) corresponds to a rotation from $O$ to $\bar{O}$.
Both the switching fields are lower compared to $\mathcal{E}_c$ of 40$\sqrt{2}$ kV/cm along
[101] direction.  In the tetragonal phase as well,
switching from $T$ to $\bar{T}$ occurs at the field of 25$\sqrt{2}$ kV/cm 
along [101] direction, as revealed by jump from point 7 to 8 in  Fig. 2c, and
at the field of 15$\sqrt{3}$ kV/cm along [111] direction as revealed by
a jump from point 5 to 6 in Fig. 2d; both $\mathcal{E}_c$'s 
are much lower than the coercive field of $\mathcal{E}_c$ = 60 kV/cm
along [001] direction. Thus, our finding for the rhombohedral phase (Fig. 1c), 
is also seen to be true for orthorhombic and tetragonal phases:       
switching with fields applied along polar direction occurs at much higher
fields than that applied along other crystallographic directions.
While the polarization rotation\cite{fu-cohen} through monoclinic phases 
(which are known to be present near the  morphotropic
phase boundaries \cite{bellaiche, noheda1, noheda2}) has been shown to be 
responsible for giant piezoelectric response, we show that it facilitates
polarization switching at lower fields and is crucial for memory applications.

We further investigate the existence and thermodynamic stability 
of monoclinic phases through $E-T$ phase diagrams. The electric field is turned on at the beginning of 
the simulations, and is kept constant. The temperature is then increased (decreased) in order 
to perform heating (cooling) simulations which give the polarization, and dielectric susceptibility
as a function of temperature. We simulated temperature dependent phase
transitions for fields along [001], [101] and [111] directions varying from 10 kV/cm
to 100 kV/cm in steps of 5 kV/cm. The contour plots of eigenvalues of 
dielectric tensor as a function of field and temperature naturally give a picture of the electric 
field - temperature ($E-T$) phase diagrams (Fig. 4). As the peak in dielectric constant marks a
transition, high density of contours visible clearly as dark regions resembling thick
lines/curves (Fig 4) mark the phase boundaries in $E-T$ phase diagrams. These lines
meet T-axis at 320 K, 250 K and 220 K at $E=0$, which are the transitions temperatures
in zero electric field\cite{PRL}. Hysteresis in heating and cooling simulations give 
small errors ($\sim$ 10 K and 10 kV/cm) in the phase boundaries .

\begin{figure}[htbp]
\centering
\includegraphics[scale=0.26,angle=-90]{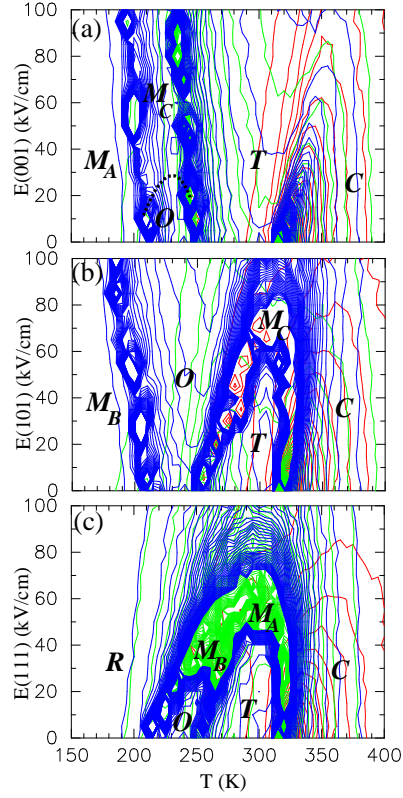}
\caption{(Color online) $E-T$ phase diagrams expressed as contour plots of
eigenvalues of the dielectric tensor. High density of contours near the peak values
of dielectric response give rise to thick and dark lines which form the boundaries
between different stable phases. A crossover from one phase to another at such a
boundary is also seen as a discontinuity in the slope of $\eta(E)$ in Fig. 3.}
\end{figure}

For field along [001] direction,
there is no qualitative change in the phase transition 
behavior with increasing field (Fig 4a);
the temperature range of the stable tetragonal phase expands with field.  
We note that a nonzero field breaks the symmetry of the system and the transition
from C to T phase becomes diffused, and above the field of 50 kV/cm, there is 
virtually no difference between the T and C phases.
The  orthorhombic phase transforms to  monoclinic $M_C$ phase above
a field of about 30 kV/cm, as characterized by a change in
the slope of strain as a function of electric field (see Fig. 3).
When the field is applied along the [101] direction, 
the temperature interval of the stable orthorhombic phase expands with field.
The tetragonal phase transforms through an intermediate monoclinic M$_C$ phase
to an orthorhombic phase above 80$\sqrt{2}$ kV/cm. Boundaries of the region of stability
of the M$_C$ phase are characterized by peaks in eigenvalues of dielectric response
appearing as thick dark lines (see Fig. 4b). 
Similarly for field along [111] direction, both tetragonal and orthorhombic phases transform 
at fields of about 40$\sqrt{3}$ and 20$\sqrt{3}$ kV/cm to $M_A$ and $M_B$ phases, respectively. 
Above the field of 60$\sqrt{3}$ kV/cm, there is no clear boundary between rhombohedral and 
cubic phases. The lines of transition from monoclinic phases to 
T, O and R phases (with increasing fields) in our $E-T$  phase diagrams are topologically
similar to those in Bell's work\cite{bell} based on phenomenological theory and even the
fields of these transitions agree within $\sim$20\% with Bell's estimates.
However, our phase diagram differs in an important way: 
the region of stable monoclinic phases appears only above nonzero electric fields
and its boundary with $T$, $O$ and $R$ phases at lower fields is characterized 
by (a) peak in the dielectric response (Fig. 4) and (b) change in the slope of 
$\eta$ changes (Fig. 3).

The sequence of phases appearing along the switching paths followed by the 
system during {\bf P-E} hysteresis can be understood from the constant temperature 
lines in the $E-T$ phase diagrams. For example, applying field along [111]
direction of the tetragonal phase at 290 K would result in {\bf P} 
tracing the path described in Fig. 2d. The connectivity between different monoclinic
phases and $T$, $O$, $R$ phases in the $E-T$ phase diagram is consistent with
that obtained from the Cohen's cube\cite{fu-cohen}. 
The change in the slope of strain as a function of field (see Fig 3) at
a crossover from a $T$ or $O$ or $R$ phase to a monoclinic one gives rise to
the giant piezoelectric response arising at finite fields in
experiments\cite{wada,park2}.

To summarize, we have used effective hamiltonian in molecular dynamics simulations 
and have shown that: (1) switching
fields are highest for the cases when the external electric field is applied along the 
direction of the polar axis, (2) switching occurs at low values of fields along
other directions because of the polarization rotation through monoclinic phases.
Stability of these monoclinic phases has been carefully analysed through determination
of the $E-T$ phase diagrams, which should be fundamental to the design of both
electromechanical and memory type of devices.

J.P. thanks CSIR and T.N. thanks JNCASR, MEXT-Japan and JSPS. We acknowledge
helpful discussions with Dhananjai Pandey and Subir K Das. We thank
the CCMS supercomputing facility at JNCASR.

\end{document}